\documentclass[aps,prl,twocolumn,groupedaddress]{revtex4}
\usepackage{graphicx}
\usepackage{xcolor}
\usepackage{float}

\usepackage{xr-hyper}

\newcommand{\etal}{\textit{et al{.}}}

\usepackage[shortlabels]{enumitem}

\newcommand{\press}[1]{\SI{#1}{\giga\pascal}}

\newcommand{\vol}[1]{\SI{#1}{\angstrom^3 / uc}}

\usepackage{siunitx} 
\usepackage{amsmath} 

\begin{document}

\title{Detectability of core level crossing and electronic topological transformations: the case of Osmium}

\author{Gavin A. Woolman and Graeme J. Ackland$^1$}
\email[]{gjackland@ed.ac.uk }

\affiliation{ $^1$Centre for Science at Extreme Conditions and School of Physics and Astronomy, University of Edinburgh, Edinburgh, U.K.}

\begin{abstract}
Osmium, the least compressible metal,  has recently been observed to undergo abrupt changes in the c/a ratio at extreme pressures. These are claimed to provide evidence for two unusual electronic behaviours: a crossing of the semicore 4f and 5p levels, and an electronic topological transition.   We demonstrate that these two electronic phenomena are readily reproduced and understood in density functional theory, but that neither perturb the trend in c/a ratio against pressure. Hence the observed anomalies in c/a must have another cause.  Osmium is also notable for its high yield stress: the c/a anomalies lie well within the differential strains which osmium can support.  We propose that observed c/a changes can arise from  mechanical yield of crystallites with strong preferred orientation under high deviatoric stress in the experimental data. We discuss what evidence remains for the more general hypothesis that core-level overlap under pressure can have measurable effects on the crystal structure in any material.
\end{abstract}
\maketitle

 Density functional theory has recently predicted that the $4f$ and $5p$ core levels overlap in high-pressure 5d transition metals.  Consequently, there has been considerable recent interest in whether such overlap has any experimentally visible signal. 

Osmium is $hcp$, meaning it has two independent lattice parameters which may show some unusual behavior.   It also has the lowest compressibility \cite{kenichi2004bulk,cynn2002osmium} and highest yield strength \cite{weinberger2008osmium} of any elemental metal. This makes it a particularly favourable candidate for studies where one ambition is reaching the highest possible pressures \cite{dubrovinsky2015most}.

Osmium also has one of the most complex electronic structures of any element.  The atomic ground state is [Xe].4f14.5p6.5d6.6s2.  Solid osmium adopts the  hexagonal close packed structure with hybridization of the $5d$ and $6s$ bands.  There is no pressure-induced structural phase transition, but there may be electronic topological transitions (ETT) if a band which does not cross the Fermi level at low pressure does so at higher pressure, or vice versa.

Abrupt changes in $c/a$ under pressure have been observed in some experiments \cite{occelli2004experimental,dubrovinsky2015most}. These have been associated with an ETT osmium undergoes at high pressures. However, other experiments across the same pressure range, observed no such changes \cite{KMCP20,PVV17}. 
It is unclear if an ETT such as that in osmium can be detected in the crystal structure. Phonon softening was reported in tantalum \cite{zhang2019pressure}, but similar effects have been discounted in other heavy elements, e.g. zinc where a combination of inelastic neutron scattering, \textit{ab initio} calculations, and application of Betteridge's Law showed that no coupling to c/a or phonons exists from an ETT \cite{klotz1998there,steinle2001absence,li2000phonon,rao2001comment}.

In addition to changes in the character of the valence electrons, the $4f$ and $5p$ core electron bands come to overlap in energy at high pressures: the more compact $4f$ orbitals are energetically favored at high pressure compared with the more extended $5p$ orbitals. This effect is common for $5d$ metals \cite{TKEJ16}, and has been associated with anomalies in the c/a of osmium. 

In this letter, we examine whether any signatures of core-level crossing and ETT in osmium are detectable in the crystal structure. 
We select osmium because it has two lattice parameters, and it was subject of a high profile recent experimental study \cite{dubrovinsky2015most}. 
We use both VASP and all-electron (Wien2k) calculations, within the formalism of density functional theory (DFT). Numerous calculations suggest that DFT is an appropriate method to study osmium, even at the highest pressures \cite{hebbache2004ab,ma2005electronic,liang2006first,dubrovinsky2015most,TKEJ16}. 


Extensive calculations were conducted for an array of unit-cells with volumes ranging between $17.0$ and \SI{29.4}{\angstrom^3} in steps of \SI{0.252}{\angstrom^3/uc}, and with $c/a$ ratios ranging from 1.565 to 1.615 in steps of 0.025. \textit{Ab initio} calculations of the total energy were conducted on this dense grid of $c/a$ and volumes. The hydrostatic equilibrium state at each volume was found by minimizing energy with respect to the $c/a$ ratio at constant volume. 

These calculations were repeated with several different exchange-and-correlation functionals: PBE, SCAN and LDA \cite{PBE96,SCAN,LDA} and two different \textit{ab initio} codes: the projector-augmented-wave pseudopotential code VASP \cite{VASP1,VASP2}; and the all-electron augmented-plane-wave with local-orbitals method as implemented by the Wien2k:v21.1 code \cite{Wien2k}. Spin-orbit coupling is essential, but otherwise the methods give small numerical differences. All sets of calculations gave qualitatively similar results. We see no reason to suppose other  treatments\cite{schimka2013lattice,ma2005electronic,ramchandani1980electronic,ren2012random,liang2006first} would be qualitatively different. Therefore, we focus on the calculations using Wien2k and the PBE approximation for exchange-and-correlation effects. The other results are included in supplementary material \cite{supplementary_material_osmium} and point to the same overall conclusions.

\begin{figure}
    \centering
    \includegraphics[width=\linewidth]{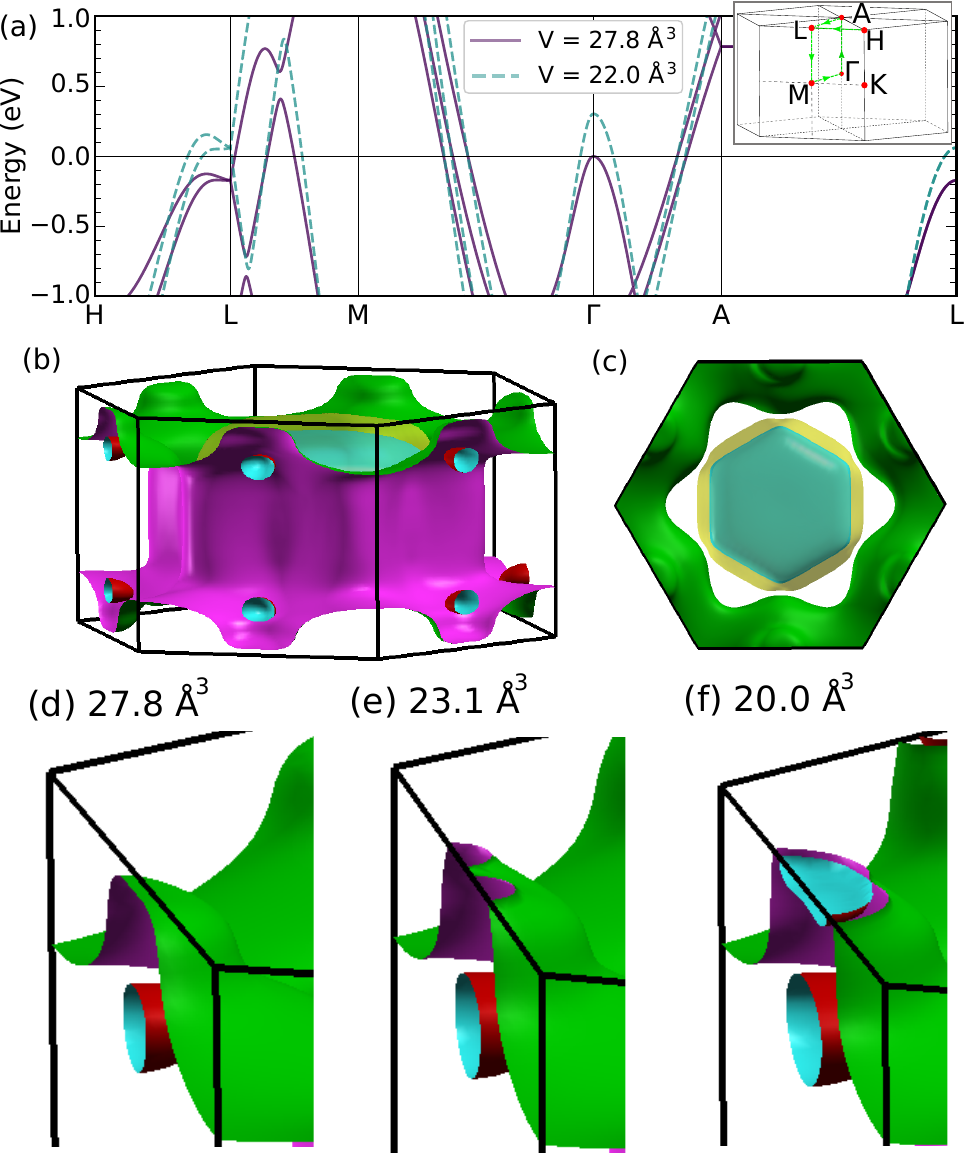}
    \caption{(a) Calculated bandstructure of osmium with a unit-cell volume of \vol{27.8} and a c/a ratio of 1.580, and \vol{20.0}, $c/a = 1.600$, respectively. These calculations used the PBE exchange-correlation potential, although the bandstructure is very similar to the SCAN and LDA calculations (see supplementary material \cite{supplementary_material_osmium}). Inset: diagram of the Brillouin-zone for the space group P6$_3$/mmc \cite{AOFT14}. (b) and (c), two different views of the Fermi surface of Osmium with $V= \vol{27.8}$, $c/a = 1.580$. (d); (e); (f): calculated Fermi surface near the L-point at \vol{27.8}, $c/a = 1.580$; \vol{23.1}, $c/a = 1.592$; and \vol{20.0}, $c/a = 1.601$, respectively.}
    \label{fig:Fermi_surfaces}
\end{figure}

Figure \ref{fig:Fermi_surfaces}(a) shows the bandstructure of osmium. Figures \ref{fig:Fermi_surfaces}(b) and (c) show the calculated Fermi surface at zero pressure. Experimental de Haas-van Alphen measures of the Osmium Fermi-surface \cite{KA70} show excellent agreement with our Wien2k calculations of the Fermi surface (see supplementary material \cite{supplementary_material_osmium}). There are two distinct qualitative changes to the bandstructure of osmium under increasing pressure. 

The first is the growth, under increasing pressure, of a hole pocket at the $\Gamma$ point. As volume is reduced, the valence-band maximum at $\Gamma$ increases. Assuming the equilibrium $c/a$ ratio, the maximum at $\Gamma$ never drops below the Fermi energy at any volume: no topological transition occurs.
However, the ETT at $\Gamma$ can be induced by shear strain, if the $c/a$ ratio is increased above the calculated hydrostatic value. For example, if experimental lattice parameters are used for a calculation with the SCAN exchange-correlation potential, an ETT at $\Gamma$ occurs. This illustrates how deciding whether to use experimental lattice parameters, or those obtained from optimizing the geometry, can qualitatively change the results of a calculation.

The second electronic-topological transition is really a series of transitions which occur in the vicinity of the L-point, $(\frac{1}{2},0,\frac{1}{2})$.  At ambient pressure, the saddle-point where several bands converge at L lies just below the Fermi surface. As volume is decreased (with an accompanying increase in c/a), a hole pocket emerges. Figures\ \ref{fig:Fermi_surfaces}(d), (e), and (f) show how the shape of the Fermi surface around L changes with reducing volume. There is an initial topological change as the maximum along the L--H line rises above the Fermi energy, connecting the large Fermi surface to its image in the second Brillouin zone. At lower volumes a further topological transition takes place, as the band-crossing at L rises above the Fermi energy and a hole pocket appears.  

\begin{figure}
    \centering
    \includegraphics[width=\linewidth]{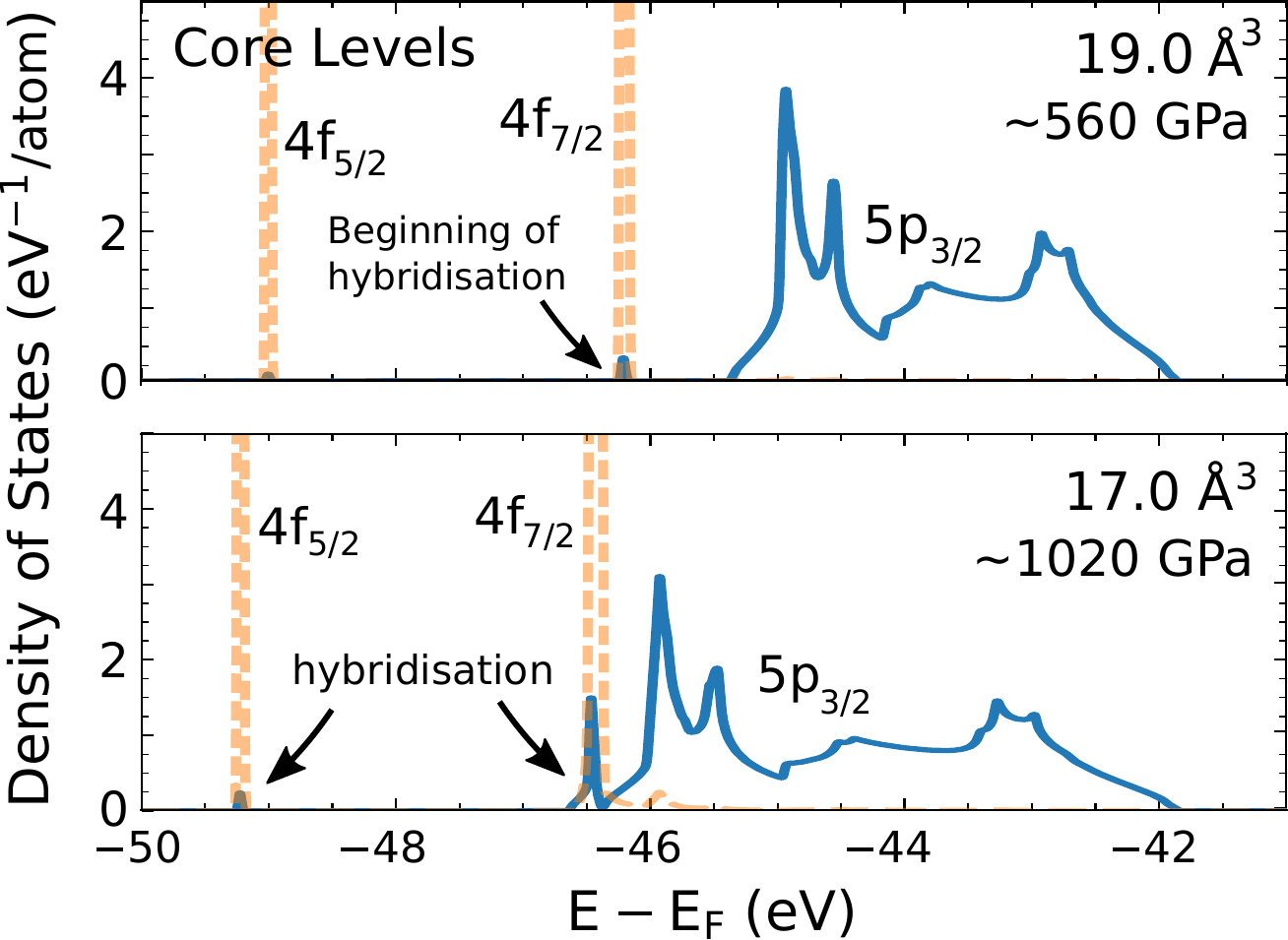}
    \caption{Projected density of states of the $4f$ and $5p_{3/2}$ core levels at two different volumes calculated using the PBE exchange-and-correlation potential: (top) at $V = \SI{19.0}{\angstrom^3 / uc}$ and $c/a = 1.6032$; (bottom) $V=\SI{17.0}{\angstrom^3 / uc}$ and $c/a = 1.6092$. }
    \label{fig:DOSplot_Combined}
\end{figure}

Figure \ref{fig:DOSplot_Combined} shows how our high pressure Wien2k calculations capture crossing of the core-level $4f_{7/2}$ and $5p_{3/2}$ orbitals well below the Fermi level. The $4f$ and $5p$ orbitals are initially distinct, separated by a gap of several electron-volts. At lower volumes / higher pressures the $5p$ band broadens and lowers in energy, eventually hybridizing with the $4f_{7/2}$ orbitals. It is this hybridisation that Dubrovinsky \etal{} claim perturbs the $c/a$ ratio of osmium. However, our calculations show that neither core-level hybridisation, nor the ETT, affect the $c/a$ ratio of osmium.

\begin{figure}
    \centering
    \includegraphics[width=\linewidth]{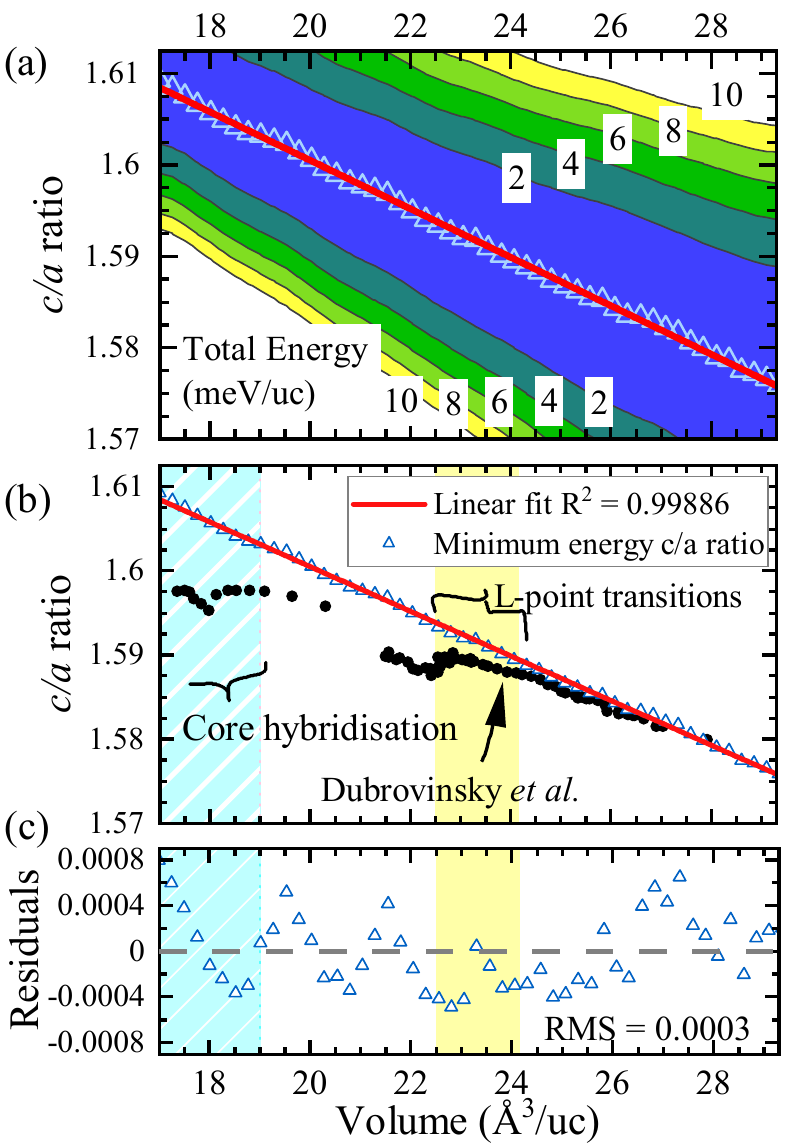}
    \caption{(a) Contour plot of energy against c/a ratio and volume, compared to the minimum energy at that volume. The  $c/a$ ratio corresponding to hydrostatic pressure  at each volume is shown with a blue triangle, and a linear fit of the hydrostatic $c/a$ ratios is also shown as a guide to the eye. The PBE exchange-correlation functional was used. Gaussian smoothing has been applied. A figure without smoothing is in supplementary material \cite{supplementary_material_osmium}. (b) Linear fit of the lowest energy $c/a$ ratio at each volume, against volume. Intervals over which the electronic-topological transition, and core-hybridisation, take place are shown (c) Residuals of the linear fit.}
    \label{fig:PBE_linfit_contours}
\end{figure}

For all sets of calculations, the $c/a$ ratio smoothly increases with decreasing volume. This smooth behavior is best illustrated by a contour plot of energy against $c/a$ ratio and volume [Fig.\ \ref{fig:PBE_linfit_contours}(a)]. There are no clear wrinkles or perturbations in the energy surface which can be associated to the ETT or core-level hybridisation.   

The hydrostatic $c/a$ ratio increases with decreasing volume. Remarkably, the trend is very well approximated by a straight line. Figure\ \ref{fig:PBE_linfit_contours}(b) shows a linear fit of the hydrostatic c/a ratio vs volume. There are no significant peaks or troughs in the residuals to indicate a deviation from linear behavior that can be associated with the L-point transition, or core hybridisation [Fig.\ \ref{fig:PBE_linfit_contours}(c)]. The root-mean-square of the residuals for the linear fit is an order of magnitude lower than the size of the anomalies in c/a seen by Dubrovinsky. The linear increase in $c/a$ ratio with decreasing volume was seen for all three choices of exchange--correlation potential and for the calculations using VASP (see supplementary material \cite{supplementary_material_osmium}). There are no significant perturbations in our calculated $c/a$ vs volume which can be associated with the calculated ETT or core-hybridization.
 
\begin{figure}[h]
    \centering
    \includegraphics[width=\linewidth]{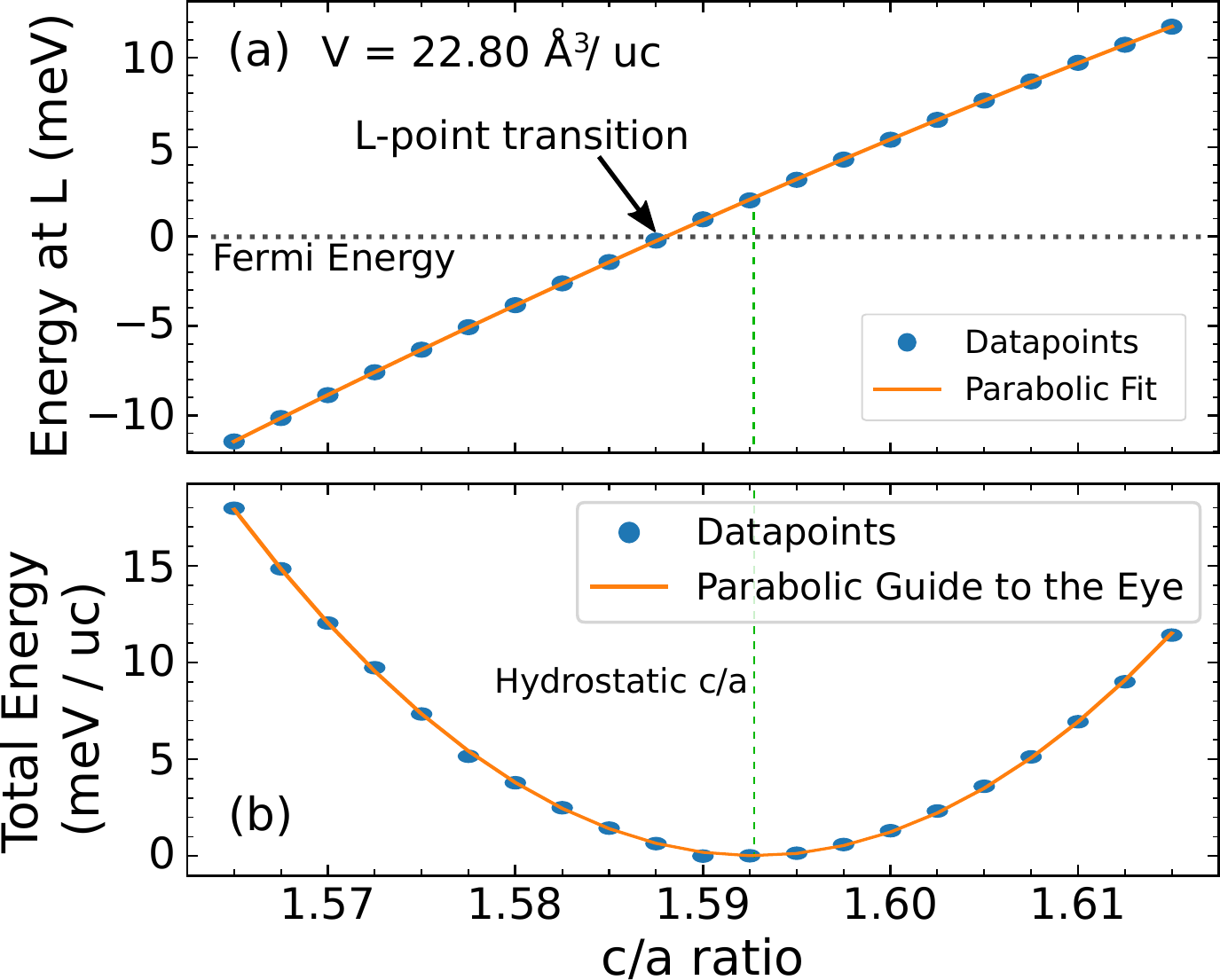}
    \caption{(a) Energy of the L-point crossing vs $c/a$ ratio at constant volume $V = \vol{22.80}$, using the PBE exchange-and-correlation potential. (b) Total energy vs $c/a$ ratio. A parabolic fit is added as a guide to the eye. }
    \label{fig:Lpoint_Energy_vs_coa_edited}
\end{figure}

As noted before, the core-state overlap and/or ETT can be driven by non-hydrostatic stress as well as pressure. An even more sensitive test of whether the core or ETT transitions are coupled to $c/a$ is to examine how the bandstructure varies with $c/a$ at fixed volume. When branches of the bandstructure at the L-point pass though the Fermi energy their energy varies continuously and smoothly with $c/a$, as does the total energy of the unit cell (Fig.\ \ref{fig:Lpoint_Energy_vs_coa_edited}). Features in the bandstructure change smoothly with decreasing volume, with no discernible perturbations at the onset of the core-level crossing. 

The broadening of the core states is driven by density, not $c/a$. Changing the $c/a$ ratio has a negligible effect on the density of states of the core levels (see supplementary material \cite{supplementary_material_osmium}). There is no discontinuous behavior arising from coupling between $c/a$ and the ETT or core-state overlap.



Third-order Birch--Murnaghan and Vinet equations of state were fitted to energy as a function of volume with the hydrostatic $c/a$ ratio. Table \ref{table:EOS fits} shows the fitting parameters. Overall there is good agreement with experiment: the discrepancy is similar to the uncertainty arising from the choice of equation of state. Fitting using the Vinet formula, rather than a third-order Birch--Murnaghan, reduces the bulk-modulus fitting parameter $B_0$ by approximately \SI{5}{\percent}. 

\begin{table}[h]
\caption{Fitted parameters for third-order Birch--Murnaghan (top) and Vinet (bottom) equations of state for this work --- calculated using LDA, PBE, and SCAN --- and for experimental work. }
\label{table:EOS fits}
\begin{tabular}{ c  c  c  c }
    ~ & $B_0$(GPa) & $B'$ & $V_0 (\si{\angstrom^3} / uc)$  \\
    \hline
    LDA & 433 & 4.5 & 27.7 \\
    PBE & 386 & 4.5 & 28.9 \\
     SCAN & 448 & 4.5 & 27.7 \\
    Dubrovinsky \etal{} \cite{dubrovinsky2015most} & 399 & 4.04 & 28.02 \\
    Kenichi \cite{K04} & 395 & 4.5 & 27.976 \\
    Pantea \etal{} \cite{PSLB09} & 405 & - & - \\
    \hline
    LDA & 417 & 5.0 & 27.7 \\
    PBE & 364 & 5.1 & 28.9 \\
    SCAN & 433 & 4.9 & 27.7 \\
    Dubrovinsky \etal{} \cite{dubrovinsky2015most} & 380 & 4.48 & 28.08 \\
    \hline
\end{tabular}
\end{table}

In their work, Dubrovinsky \etal{} \cite{dubrovinsky2015most} compared Birch--Murnaghan equations of state which they had fitted to subsets of their data from different pressure ranges. They found that the fit made to data at pressures above \press{400} had a significantly smaller $B_0$ parameter, when compared to the fits made on data at pressures below \press{400}. They claimed \textit{``these experimentally observed peculiarities are not artefacts and require an explanation"}.

This led us to investigate the sensitivity of results to the details of the fitting procedure.  To test this,
we generated ``pressure--volume data" using the Lennard--Jones function, which has a simple analytic form free from anomalies.
We then fitted subsets of this smooth data using a third-order Birch-Murnaghan equation.  The subsets exhibited dramatic differences in the fitted parameters, even though the underlying `true' equation of state had no discontinuities. 
This illustrates that obtaining different fitting parameters from different subsets of pressure--volume data is not good evidence of a transition. 
Although the peculiarities observed by Dubrovinsky \etal{} may not be an artefact of the experimental data, it could be an artefact of the fitting process. By applying the same flawed analysis to our calculated data for osmium, we are also able to generate spurious discontinuities.


Although the discontinuity in equation of state was a fitting artefact, the experimental evidence for anomalous changes in $c/a$ under pressure is convincing \cite{dubrovinsky2015most}. However, our calculations show that these transitions have no coupling to the $c/a$ ratio: another explanation for the observations is needed. 

One possible cause of the anomalies in $c/a$ is non-hydrostatic stress. Wienberger \etal{} \cite{weinberger2008osmium} have shown that Osmium is able to sustain large anisotropic stresses before yielding: differential stress on the order of \press{10} at a pressure of \press{26}. 
They also showed that, under the large differential stresses which can be achieved in a diamond-anvil cell, osmium can support deviatoric strain in its $c/a$ ratio of up to $0.004$ away from the hydrostatic value.  X-ray data \cite{dubrovinsky2015most,weinberger2008osmium} showed that polycrystalline Osmium has a strong preferred orientation when under non-hydrostatic pressure in a diamond-anvil cell: consequently, one can expect the deviatoric stress to be aligned along the $c$-axis in all crystallites, leading primarily to a shift the the diffraction peaks rather than simple strain-broadening. Wienberger \etal{} show that large differential stress can be sustained by osmium at high pressure, leading to large differences in the measured $c/a$ from what would be achieved under hydrostatic conditions. The apparent $c/a$ under pressure would likely show anomalies if the sample yielded to plastic deformation. It is also worth noting that Wienberger \etal{} showed deviations in $c/a$ of $0.004$ at just \press{26}, an order of magnitude lower pressure than that reached by Dubrovinsky \etal{}; it is likely that, at greater pressures, larger deviations in $c/a$ could be sustained before yielding.

\begin{figure}
    \centering
    \includegraphics[width=\linewidth]{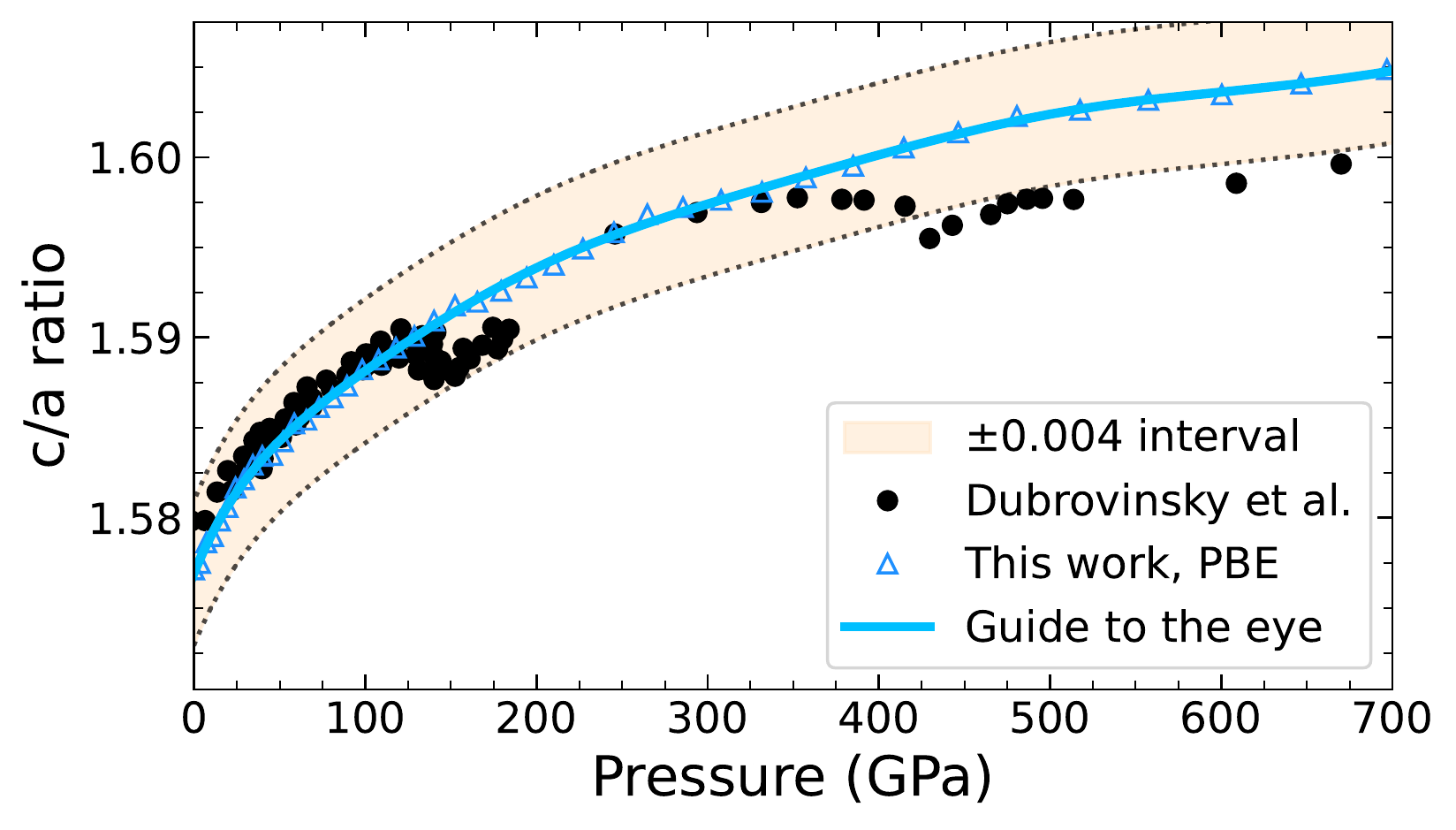}
    \caption{Calculated hydrostatic $c/a$ ratio as a function of pressure. An interval of $\pm 0.004$ is shown, informed by the deviations in apparent $c/a$ ratio observed by Weinberger \etal{} \cite{weinberger2008osmium}. When comparing with experiment, we note that there are issues with the accurate determination of pressure such that matching pressures between experiment and calculation may be problematic.}
    \label{fig:Wienberger}
\end{figure}

Conservatively taking $0.004$ as the deviation in $c/a$ which can be achieved before plastic yield, an interval of possible $c/a$ ratios can be drawn around the calculated hydrostatic value. Figure\ \ref{fig:Wienberger} shows that the anomalies in the $c/a$ vs pressure curve observed \cite{dubrovinsky2015most} at 150 and \press{440}, are well within this $\pm 0.004$ interval. 

Thus, Fig.\ \ref{fig:Wienberger} shows that kinks in the $c/a$ ratio in a diamond-anvil cell experiment could be explained by a yield process:  anisotropic stress built up in compression of osmium, the least compressible metal, reaches the yield stress and relaxes towards hydrostatic conditions. In this scenario, the correlation with first ETT and then core-overlap is just coincidence. Yielding is an irreversible process, so this could in principle be tested by measuring the $c/a$ ratio upon decompression.

It is possible that the electronic topological transition at $L$, or the core-level overlap, could have some subtle effect on things like the yield stress.  Compression of similar samples in the same uniaxial geometry may even make the result reproducible. However, the authors know of no evidence supporting this hypothesis, and later experimental work by Perreault \etal{} \cite{PVV17} and Kuzovnikov \etal{} \cite{KMCP20} did not reproduce the anomalies seen by Dubrovinsky \etal{}, which further indicates that the proximity of the ETT and core-overlap to the anomalies is coincidental. 

To summarize: A thorough and comprehensive set of \textit{ab initio} calculations were performed over a range of $c/a$ ratios and volumes. Electronic topological transitions at $L$ were observed, occurring at volumes between $22.5$ and \SI{24.8}{\angstrom^3/uc}, the uncertainty being determined from using different exchange and correlation potentials.   
The $5p_{3/2}$ and $4f_{7/2}$ core-levels hybridize as volume decreases below \SI{19}{\angstrom^3/uc}. 

However, these interesting electronic phenomena have no discernible effect on the equation of state, nor on the $c/a$ ratio as a function of pressure or volume, nor on the energy vs c/a and volume landscape. Indeed, to an excellent approximation the $c/a$ ratio increases linearly with decreasing volume. 

We demonstrated that the equation of state peculiarities reported by Dubrovinsky \etal{} \cite{dubrovinsky2015most} could be artefacts of the fitting process. Our calculated data is smooth, but by fitting Burch-Murnaghan equations to subsets of the data we generate a spurious anomaly, analogous to that reported experimentally.
The anomalies in $c/a$ vs pressure seen experimentally are real, and can be explained by high anisotropic stresses in the sample, which build up and eventually yield. 
The core-effect hypothesis requires one to believe that DFT gives an accurate calculation of core state overlap, while at the same time is qualitatively wrong about the effects of core states on structure.

However, it seems more plausible that the proximity in pressure between experimentally observed anomalies and the calculated ETT and core-level hybridisation, is coincidental.

Our calculations here considered Osmium, but we expect them to apply to all $5d$ transition metals.  We have found no mechanism by which subtle electronic effects such as ETT or core overlap can cause detectable structural anomalies in any material. 

The only evidence the authors know of that core-level overlap has any effect is based on the proximity of a DFT-calculated overlap pressure to the pressure of experimentally-measured anomalies. However, with the absence of any theoretical work showing a link between core-level overlap and the crystal structure, and with the presence of experimental work indicating that these anomalies can more plausibly be explained by anisotropic stress in the diamond-anvil cell \cite{weinberger2008osmium}, it is prudent to question whether core-level overlap has a measurable impact on the crystal structure in any material.

\bibliographystyle{apsrev}
\bibliography{HeavyMetal}

\subsection*{Acknowledgements}
The authors would like to acknowledge the support of the European Research Council (ERC) Grant "Hecate" reference No. 695527, and EPSRC for the UKCP grant P022561. GJA acknowledges a Royal Society Wolfson fellowship. This work used the ARCHER2 UK National Supercomputing Service (https://www.archer2.ac.uk). For the purpose of open access, the author has applied a Creative Commons Attribution (CC BY) licence to any Author Accepted Manuscript version arising from this submission.

\end{document}